\documentclass[review,preprint,12pt]{elsarticle}

\usepackage{amsmath,amssymb}
\usepackage{graphicx}
\usepackage{geometry,natbib}
\biboptions{numbers,sort&compress}
\usepackage[breaklinks,colorlinks=true,citecolor={blue}, linkcolor= {blue}, urlcolor ={blue}]{hyperref}

\begin{document}
\begin{frontmatter}

\title{\textbf{Dielectric Substrate Dependence of Thermoelectric Transport in BLG–GaAs–BLG Heterostructures}}

\author[inst1,inst2]{Vo Van Tai}\ead{vovantai@vlu.edu.vn}
\author[inst3]{Truong Van Tuan}
\author[inst3]{Tran Trong Tai}

\author[l1,l2]{Le Tri Dat\corref{cor1}}

\author[inst1,inst2]{Nguyen Duy Vy}

\address[inst1]{Laboratory of Applied Physics, Science and Technology Advanced Institute, Van Lang University, Ho Chi Minh City, Vietnam}
\address[inst2]{Faculty of Applied Technology, School of Technology, Van Lang University, Ho Chi Minh City, Vietnam}
\address[inst3]{Tran Dai Nghia University, 189 Nguyen Oanh Str., Go Vap Dist., Ho Chi Minh City, Vietnam}

\address[l1]{Engineering Research Group, Dong Nai Technology University, Bien Hoa City, Vietnam}
\address[l2]{Faculty of Engineering, Dong Nai Technology University, Bien Hoa City, Vietnam.} \cortext[cor1]{Corresponding author: letridat@dntu.edu.vn}

\begin{abstract}
We theoretically study the thermoelectric transport 
$S$ in a double-layer bilayer graphene (BLG-GaAs-BLG) system on dielectric substrates (h-BN, Al$_2$O$_3$, HfO$_2$). Electrons interact with GaAs acoustic phonons via both the deformation potential (acDP) and piezoelectric (acPE) scattering. Results show that piezoelectric scattering dominates the total transport, especially at low carrier density and high dielectric constant. Substrate dielectric constant significantly influences 
thermopower $S$, and the thermopower of the materials is in the order of HfO$_2>$ Al$_2$O$_3>$ h-BN. When densities on two BLG layers are unequal, the contribution from acDP scattering $S_d$  decreases (increases) at low (high) densities versus equal densities, while acPE scattering $S_g$ remains stable, making 
$S$ largely $S_g$-dependent. Increasing interlayer distance $d$ enhances $S$, while higher temperature boosts $S_d$ (notably at low densities) with minimal effect on $S_g$. These insights and substrate-dependent trends demonstrate substrate engineering as a key parameter for optimizing BLG thermoelectric devices.
\end{abstract}


\end{frontmatter}

\section{INTRODUCTION}
In recent decades, both experimental and theoretical studies have intensively explored the physical properties of graphene and its related structures \cite{principi2012, Cruz2015, VanMen2020, Linh2020, Gamucci2014, Bernazzani2023}. Bilayer graphene (BLG), formed by stacking two monolayers of graphene, exhibits a distinct electronic band structure compared to monolayer graphene, characterized by quadratic dispersion and a finite effective mass, similar to that of a conventional two-dimensional electron gas (q2DEG). Among the various properties of BLG and its double-layer counterparts, the thermoelectric coefficient $S$ has attracted particular interest. Notably, the phonon-drag component of the Seebeck coefficient, $S_g$, was investigated in monolayer BLG by Kubakaddi and Bhargavi \cite{kubakaddi2010}, who considered intralayer deformation potential (acDP) scattering without including screening effects. Ansari and Ashraf \cite{ansari2017,ansari2021} extended the analysis by incorporating screening, though they neglected its temperature dependence.

For double-layer systems, Smith \cite{smith2003} and Vazifehshenas \cite{vazifehshenas2015}, among others, examined $S$ in q2DEG–q2DEG configurations with screening. More recently, studies have addressed screening effects on $S_g$ and the diffusion component $S_d$ in BLG–Air–BLG and BLG–q2DEG heterostructures \cite{tuan2023stdj, Tuan2024, Tuan2025, tuan2025hcmue}. However, to the best of our knowledge, no comprehensive investigation has yet focused on the thermoelectric properties of a BLG–GaAs–BLG double-layer system, where the two BLG layers are separated by a GaAs semiconductor and supported on dielectric substrates. In such systems, electron–phonon interactions include not only acDP but also piezoelectric (acPE) scattering arising from the GaAs layer. Moreover, the dielectric environment and interlayer screening can significantly influence transport behavior.

In this work, we present a theoretical analysis of the thermoelectric coefficient $S$ in BLG–GaAs–BLG systems under various dielectric substrates. We study the roles of interlayer screening and substrate dielectric properties, and explore the dependence of $S$ on temperature, interlayer separation $d$, and the carrier density in each BLG layer—both for equal and unequal density distributions between the layers. This study not only clarifies the dominant scattering mechanisms that govern thermoelectric transport in double-layer BLG systems but also reveals how substrate selection and layer asymmetry can be used to tailor thermoelectric performance. Our findings provide practical guidance for optimizing device parameters in experimental setups, especially in nanoscale thermoelectric devices where interfacial and dielectric effects are pronounced. Ultimately, the results presented here may serve as a theoretical foundation for the experimental realization and engineering of high-performance graphene-based thermoelectric devices, particularly in hybrid semiconductor–graphene architectures.
\begin{figure}
    \centering
    \includegraphics[width=0.6\linewidth]{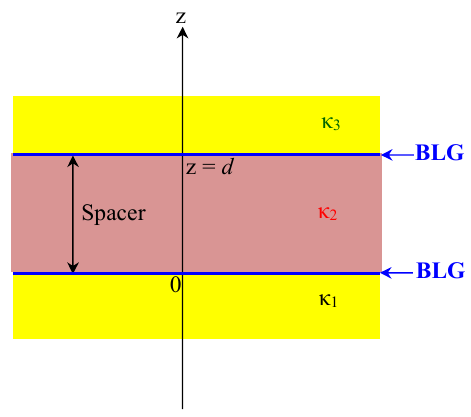}
    \caption{Schematic of the BLG–GaAs–BLG double-layer structure with different substrate dielectrics.}
    \label{fig:Figure_1}
\end{figure}

\section{METHOD}
Bilayer graphene (BLG) consists of two stacked graphene monolayers, with a thickness corresponding to two atomic layers \cite{McCann2011}. In the vicinity of the K points, BLG behaves as a gapless semimetal with a parabolic energy dispersion given by
\begin{equation}
E_{s\mathbf{k}} = \frac{s\hbar^2 k^2}{2m^*}
\end{equation}

The electronic properties of carriers in the low-energy regime can be described by an effective massive-particle Hamiltonian with a Dirac-like wave function,
\begin{equation}
\psi_{s\mathbf{k}} = \frac{1}{\sqrt{2}} \begin{pmatrix}
e^{-i2\theta_{\mathbf{k}}} \\
s
\end{pmatrix}
\end{equation}
where $s = +1$ and $-1$ correspond to the conduction and valence bands, respectively.

One of the key characteristics of a two-dimensional electron gas is its response to electromagnetic fields. The simplest approximation to describe the response of a system at short wavelengths is the self-consistent field approximation, in which each electron is assumed to move under the influence of both the external field and the induced field from all other electrons in the system \cite{DasSarma2011, Lv2010}. The dielectric function, which characterizes the screening effect within the random phase approximation (RPA), is given by \cite{Lv2010}:
\begin{equation}
\varepsilon(q,T) = 1 + \frac{2\pi e^2}{\bar{\kappa} q} \Pi(q,T)
\end{equation}
where $\Pi(q, T)$ is the temperature-dependent polarization function \cite{ansari2021}:
\begin{equation}
\label{eq:polarization}
\begin{aligned}
\Pi_{\mathrm{BLG}}(q, T) &= \frac{g_s g_v m^*}{2\pi \hbar^2} \int_0^\infty dk\, k^3 \Big\{ 
\sqrt{4k^4 + q^4} - k^2 - |k^2 - q^2| \\
&\quad + \left[ f(E_k) + f(E_k + 2\zeta) \right] \cdot 
\Big[ 2k^2 - \sqrt{4k^4 + q^4} 
+ \frac{(2k^2 - q^2)^2}{q \sqrt{q^2 - 4k^2}}\, \theta(q - 2k) \Big] 
\Big\}
\end{aligned}
\end{equation}

Here, $\zeta$ represents the chemical potential:
\begin{equation}
\zeta_{\mathrm{BLG}} = E_F
\end{equation}

When a temperature gradient $\nabla T$ is applied, charge carriers and phonons diffuse within the material, giving rise to electric and heat currents. The interplay between these phenomena is known as the thermoelectric effect. The thermopower (or Seebeck coefficient) is a quantity that characterizes this effect \cite{tuan2023stdj,Cantrell1987,cantrell1987b}. At low temperatures, the temperature gradient $\nabla T$ generates a phonon momentum flow, which drags charge carriers via the electron–phonon interaction, leading to the phonon-drag contribution to the Seebeck coefficient, denoted as $S_g$.

The BLG--GaAs--BLG double-layer structure consists of two parallel BLG layers separated by a distance $d$, with GaAs as the intervening dielectric material, as illustrated in Fig. \ref{fig:Figure_1}.

The total thermopower $S$ of the coupled bilayer system is given by \cite{tuan2023stdj,Tuan2024, Hasanov2016, Nicholas1985}:
\begin{equation}
S = \frac{\sigma_u S_u + \sigma_l S_l}{\sigma_u + \sigma_l}
\label{eq:S_total}
\end{equation}
where $\sigma$ is the electrical conductivity, expressed as $\sigma = N_s e^2 \tau_t(E_k) / m^*$, and the subscripts $u$ and $l$ correspond to the upper and lower layers, respectively.

In this work, we consider the BLG--GaAs--BLG double-layer structure placed on various dielectric substrates, with dielectric constants $\kappa_1 = \kappa_3$ and $\kappa_2 = \kappa_{\text{GaAs}}$. In this case, the phonon-drag Seebeck coefficient $S_g$ and the diffusion thermopower $S_d$, which include contributions from both deformation potential and piezoelectric interactions, are given by \cite{ansari2017,ansari2021}:
\begin{align}
S_{{g}}^\text{DP} &= - \frac{m^{*3/2} D^2 l_p}{2\sqrt{2} N_s e k_B T^2 \rho \pi^2 \hbar^3 v_s^3} 
\int_0^\infty dq \, (\hbar \omega_q)^3 \nonumber\\
&\times \int_\gamma^\infty dE_k \, G(E_k, \omega_q) \left( \frac{W_{ii}(q,T)}{V_{ii}(q)} \right)^2 
N_q \frac{f_0(E_k)[1 - f_0(E_k + \hbar \omega_q)]}{\sqrt{E_k - \gamma}},
\label{eq:Sg_DP} \\
S_{{g}}^\text{PE} &= - \frac{m^{*3/2} C_{\text{PE}}^2 D_{\text{PE}}^2 l_p}{8\sqrt{2} N_s e k_B T^2 \rho_{\text{GaAs}} \pi^3 \hbar^3 v_{\text{PE}}^2} 
\int_0^\infty dq \, (\hbar \omega_{\text{PE}})^2 \nonumber\\
&\times \int_\gamma^\infty dE_k \, G(E_k, \omega_q) \left( \frac{W_{ii}(q,T)}{V_{ii}(q)} \right)^2 
N_q \frac{f_0(E_k)[1 - f_0(E_k + \hbar \omega_q)]}{\sqrt{E_k - \gamma}}.
\label{eq:Sg_PE}
\end{align}

The diffusion thermopower is expressed as:
\begin{equation}
S_d = -\frac{1}{eT} \left[ -E_f + \frac{\langle E_k \tau_t(E_k) \rangle}{\langle \tau_t(E_k) \rangle} \right],
\label{eq:Sd}
\end{equation}
where the average relaxation time is defined as:
\begin{equation}
\langle \tau(E_k) \rangle = \frac{\int_0^{+\infty} E_k \left( \frac{\partial f_0(E_k)}{\partial E_k} \right) \tau(E_k) \, dE_k}{\int_0^{+\infty} E_k \left( \frac{\partial f_0(E_k)}{\partial E_k} \right) \, dE_k}.
\label{eq:tau_avg}
\end{equation}
Here, $E_f$ is the Fermi energy \cite{Cantrell1987,cantrell1987b}.

The screened potential $W_{ii}(q, T)$ for the double-layer system replaces the single-layer dielectric function $1/\varepsilon(q, T)$ \cite{tuan2023stdj, Tuan2024, Bhargavi2013, Hosono2014, Tuan2025,tuan2025hcmue}, and is given by:
\begin{equation}
W_{ii}(q, T) = \frac{V_{ii}(q) + [V_{ii}(q)V_{jj}(q) - V_{ij}^2(q)] \Pi_{jj}(q,T)}{[1 + V_{ii}(q)\Pi_{ii}(q,T)][1 + V_{jj}(q)\Pi_{jj}(q,T)] - V_{ij}^2(q) \Pi_{ii}(q,T)\Pi_{jj}(q,T)}.
\label{eq:Wii}
\end{equation}
The Coulomb interaction potential is:
\begin{equation}
V_{ij}(q) = \frac{2\pi e^2}{q} f_{ij}(q).
\label{eq:Vij}
\end{equation}
The temperature-dependent polarization function $\Pi_{ii}(q, T)$ for layer $i$ is given by Eq. \eqref{eq:polarization}. For the double-layer system illustrated in Fig. \ref{fig:Figure_1}, the form factors $f_{ij}(q)$ are expressed as:
\begin{align}
f_{11}(q_1) &= \frac{2(\kappa_2 \cosh(q_1 d) + \kappa_3 \sinh(q_1 d))}{\kappa_2 (\kappa_1 + \kappa_3) \cosh(q_1 d) + (\kappa_1 \kappa_3 + \kappa_2^2) \sinh(q_1 d)} \label{eq:f11} \\
f_{22}(q_2) &= \frac{2(\kappa_2 \cosh(q_2 d) + \kappa_1 \sinh(q_2 d))}{\kappa_2 (\kappa_1 + \kappa_3) \cosh(q_2 d) + (\kappa_1 \kappa_3 + \kappa_2^2) \sinh(q_2 d)} \label{eq:f22} \\
f_{12}(q_1) &= \frac{2\kappa_2}{\kappa_2 (\kappa_1 + \kappa_3) \cosh(q_1 d) + (\kappa_1 \kappa_3 + \kappa_2^2) \sinh(q_1 d)} \label{eq:f12} \\
f_{21}(q_2) &= \frac{2\kappa_2}{\kappa_2 (\kappa_1 + \kappa_3) \cosh(q_2 d) + (\kappa_1 \kappa_3 + \kappa_2^2) \sinh(q_2 d)} \label{eq:f21}
\end{align}

\section{RESULTS AND DISCUSSION}

We investigate the thermopower $S$ using the following parameters for bilayer graphene (BLG): $m^* = 0.033 m_e$, $D = 20~\mathrm{eV}$, $\rho = 7.6 \times 10^{-8}~\mathrm{g/cm^2}$, $v_s = 2 \times 10^6~\mathrm{cm/s}$, $l_{1p} = l_{2p} = 10~\mu\mathrm{m}$ \cite{kubakaddi2010}. For GaAs \cite{VanTan2019}: $\rho_\mathrm{GaAs} = 5.31~\mathrm{g/cm^3}$, $D_\mathrm{PE} = 2.4 \times 10^7~\mathrm{eV/cm}$, $C_\mathrm{PE} = 4.9$, $v_\mathrm{PE} = 2.7 \times 10^5~\mathrm{cm/s}$, $\kappa_\mathrm{GaAs} = 12.91$ \cite{ansari2021}. 
The background dielectric constants are taken as: $\kappa_1 = \kappa_3 = \kappa_{\mathrm{h\text{-}BN}} = 4$, $\kappa_1 = \kappa_3 = \kappa_{\mathrm{Al_2O_3}} = 12.53$, and $\kappa_1 = \kappa_3 = \kappa_{\mathrm{HfO_2}} = 22$ \cite{Wilk,kumar2022}. 

\begin{figure}[!h] \centering
\includegraphics[width=1\linewidth]{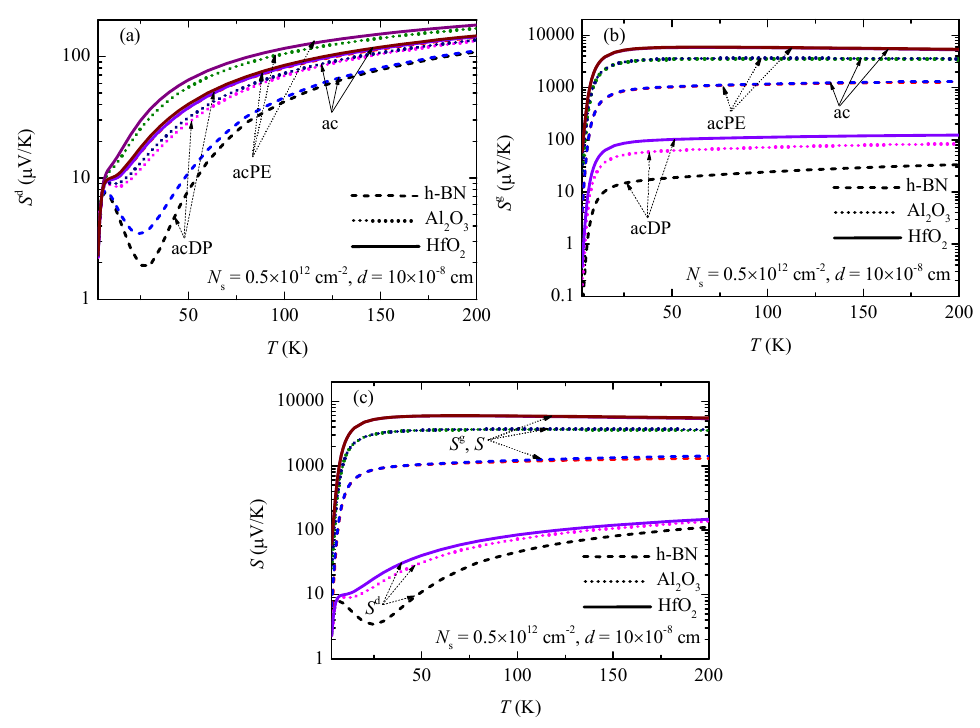}
\caption{Variation of the thermopower $S$ as a function of temperature at a carrier density of $N_s = 0.5 \times 10^{12}~\mathrm{cm^{-2}}$ and interlayer distance $d = 10 \times 10^{-8}~\mathrm{cm}$, with different background dielectric materials: (a) Diffusive thermopower component $S_d$ due to acoustic deformation potential (acDP), piezoelectric potential (acPE), and acoustic phonon (ac) scattering; (b) Phonon drag thermopower component $S_g$ due to acDP, acPE, and ac phonons; (c) Total thermopower $S$ due to acoustic phonon scattering.}
    \label{fig:Figure_2}
\end{figure}

\begin{figure}[!h] \centering
\includegraphics[width=1\linewidth]{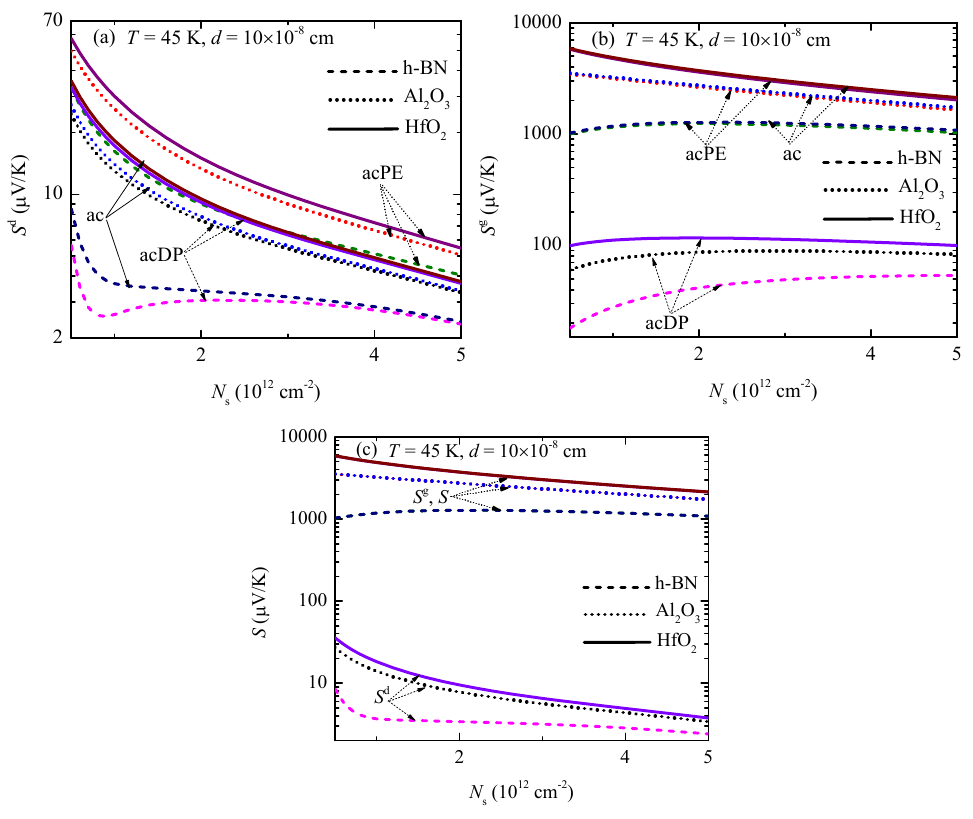}
\caption{Dependence of the thermoelectric power $S$ on carrier density at $T = 45~\mathrm{K}$ and interlayer spacing $d = 10 \times 10^{-8}~\mathrm{cm}$ for various dielectric substrates:
 (a) Diffusion thermopower $S_d$ contributed by acoustic deformation potential (acDP), piezoelectric potential (acPE), and acoustic phonon scattering;
 (b) Phonon drag thermopower $S_g$ arising from acDP, acPE, and acoustic phonons;
    (c) Overall thermopower $S$ resulting from acoustic phonon interactions.}
    \label{fig:Figure_3}
\end{figure}

\begin{figure}[!ht]\centering   \includegraphics[width=1\linewidth]{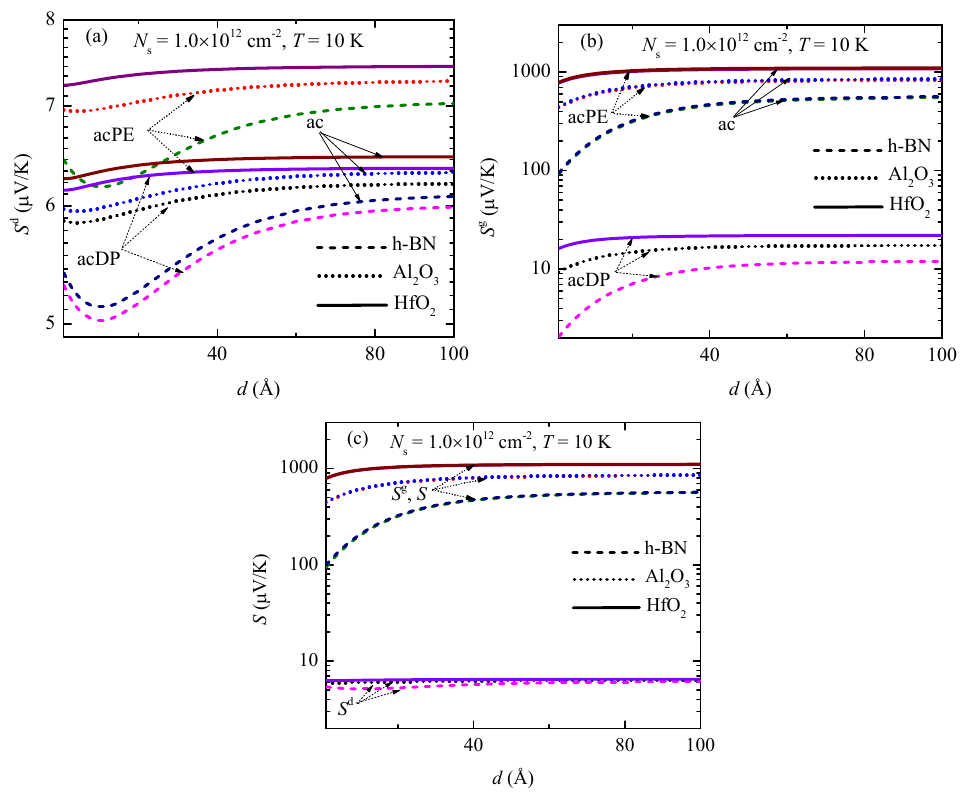}
    \caption{Dependence of the thermoelectric coefficient $S$ on the interlayer spacing $d$ at $T = 10~\mathrm{K}$ and $N_s = 10^{12}~\mathrm{cm^{-2}}$ for various dielectric environments: 
(a) Diffusion component $S_d$ contributed by acoustic deformation potential (acDP), piezoelectric (acPE), and total acoustic phonons; 
(b) Phonon drag component $S_g$ for the same scattering mechanisms; 
(c) Overall thermoelectric coefficient $S$ considering acoustic phonon contributions.}
    \label{fig:Figure_4}
\end{figure}

\begin{figure}[!h] \centering\includegraphics[width=1\linewidth]{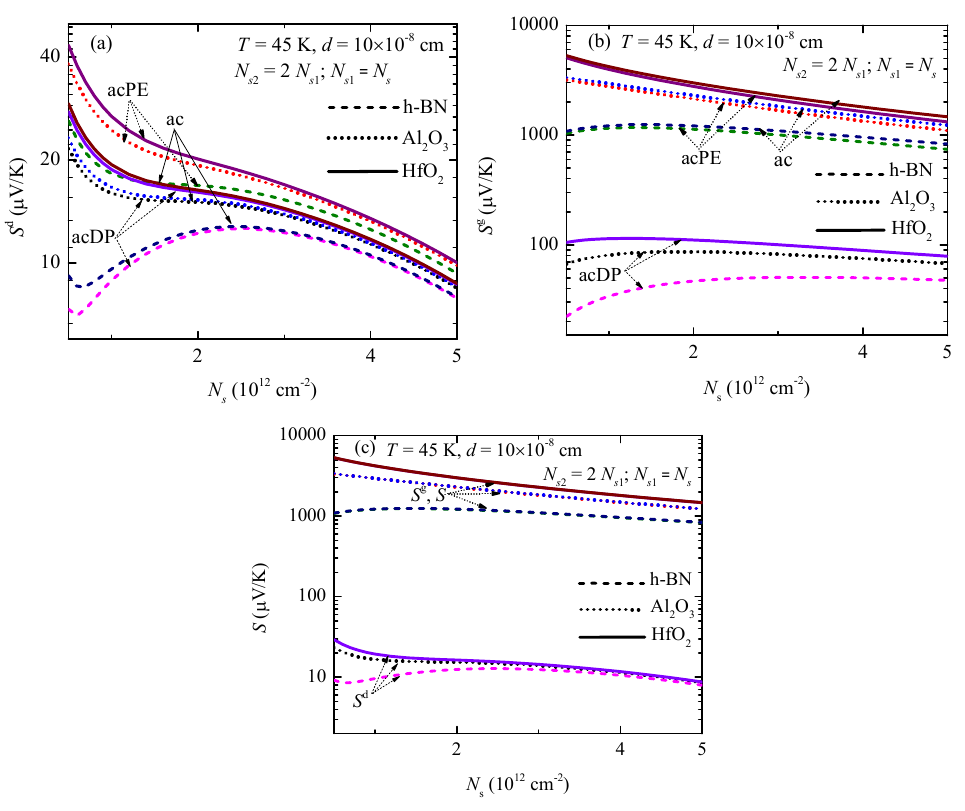}
\caption{Variation of the thermoelectric coefficient $S$ as a function of carrier density with $N_{s2} = 2 \times N_{s1}$, $d = 10 \times 10^{-8}~\mathrm{cm}$, $T = 45~\mathrm{K}$ for different dielectric substrates: 
(a) Diffusion thermopower $S_d$ due to acoustic deformation potential (acDP), piezoelectric interaction (acPE), and total acoustic phonons (ac); 
(b) Phonon drag thermopower $S_g$ due to acDP, acPE, and ac phonons; and 
(c) Total thermoelectric coefficient $S$ due to acoustic phonons.}
    \label{fig:Figure_5}
\end{figure}

\begin{figure}[!h] \centering
\includegraphics[width=1\linewidth]{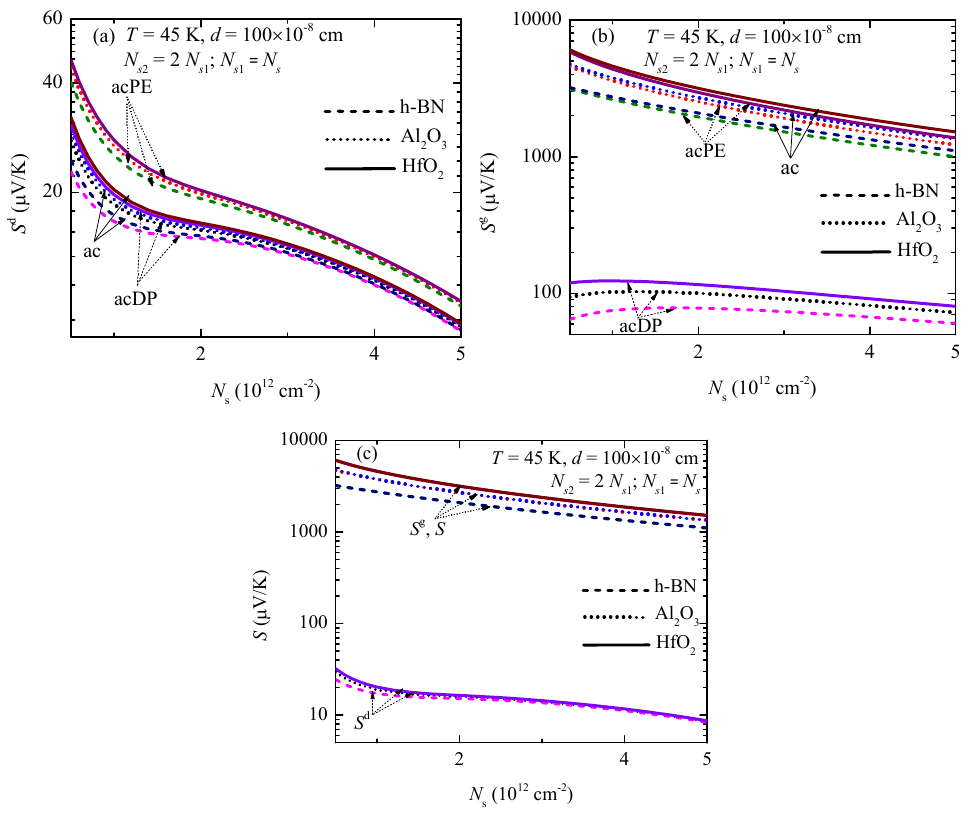}
\caption{Thermoelectric coefficient $S$ versus carrier density with $N_{s2} = 2 \times N_{s1}$, $d = 100 \times 10^{-8}~\mathrm{cm}$, and $T = 45~\mathrm{K}$ for various dielectric substrates: 
(a) Diffusion thermopower $S_d$ from acoustic deformation potential (acDP), piezoelectric (acPE), and total acoustic (ac) scattering; 
(b) Phonon drag thermopower $S_g$ for the same mechanisms; 
(c) Total thermoelectric coefficient $S$ due to acoustic phonons.}
    \label{fig:Figure_6}
\end{figure}

\begin{figure} [!h]\centering
\includegraphics[width=1\linewidth]{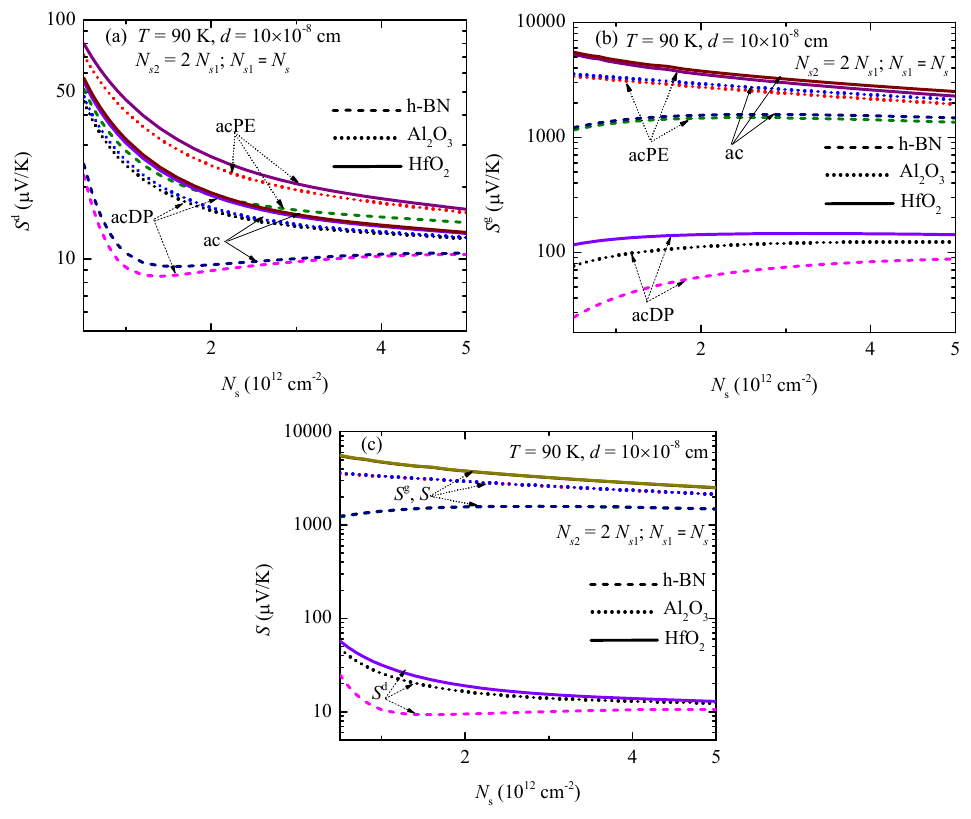}
\caption{Variation of the thermoelectric coefficient $S$ with carrier density for $N_{\mathrm{s2}} = 2 \times N_{\mathrm{s1}}$, $d = 10 \times 10^{-8}~\mathrm{cm}$, and $T = 90~\mathrm{K}$ under different dielectric substrates: 
(a) Diffusive thermopower $S_{\mathrm{d}}$ due to acoustic deformation potential (acDP), piezoelectric potential (acPE), and total acoustic phonons (ac); 
(b) Phonon drag thermopower $S_{\mathrm{g}}$ due to acDP, acPE, and ac; and 
(c) Total thermopower $S$ due to acoustic phonons.}
    \label{fig:Figure_7}
\end{figure}

In Figs. \ref{fig:Figure_2} to \ref{fig:Figure_4}, we analyze the case of symmetric double layers. The case of asymmetric carrier densities between the two BLG layers is investigated and presented in Figs. \ref{fig:Figure_5} to \ref{fig:Figure_7}.

Fig. \ref{fig:Figure_2} illustrates the temperature dependence of the diffusive thermopower $S_d$, phonon drag thermopower $S_g$, and total thermopower $S$ at a carrier density $N_s = 0.5 \times 10^{12}~\mathrm{cm^{-2}}$ and interlayer spacing $d = 10 \times 10^{-8}~\mathrm{cm}$ for different background dielectric substrates, considering acoustic phonon deformation potential (acDP), piezoelectric potential (acPE), and acoustic phonon (ac) scattering.

From Fig. \ref{fig:Figure_2}(a), $S_d$ is larger for substrates with higher dielectric constants and increases with temperature, similar to $S_d$ observed in other semiconductors \cite{Cantrell1987, cantrell1987b, Hasanov2016}, reaching the minimum value for the substrate with the lowest dielectric constant (h-BN). Among the $S_d$ contributions, scattering by acDP dominates over acPE scattering for acoustic phonons.

In contrast, Fig. \ref{fig:Figure_2}(b) shows that $S_g$ increases with temperature and saturates at high temperatures as reported in \cite{Tuan2024, Scharf2012}, with higher values for substrates with larger dielectric constants. Unlike $S_d$, acPE scattering is the primary contributor to $S_g$ compared to acDP scattering.

Figure \ref{fig:Figure_2}(c) depicts the combined contributions of $S_d$ and $S_g$ to the total thermopower $S$ via acoustic phonon scattering. The results indicate that the total thermopower $S$ is predominantly determined by the phonon drag component $S_g$. Based on the properties of the dielectric substrates such as h-BN \cite{pakdel2014, yang2012}, Al$_2$O$_3$ \cite{levin1998}, and HfO$_2$ \cite{Wilk, kumar2022}, the thermopower $S$ increases in magnitude with increasing background dielectric constant.

Figure \ref{fig:Figure_3} illustrates the variation of the thermoelectric power $S$ with carrier density at temperature $T = 45~\mathrm{K}$ and interlayer distance $d = 10 \times 10^{-8}~\mathrm{cm}$ on different dielectric substrates. From Fig. \ref{fig:Figure_3}(a), it is observed that the diffusive thermopower $S_d$ decreases with increasing carrier density, consistent with previous results \cite{Cantrell1987, cantrell1987b}, and the acoustic deformation potential (acDP) scattering dominates over piezoelectric (acPE) scattering in contributing to acoustic phonon scattering. For the phonon drag thermopower $S_g$ shown in Fig. \ref{fig:Figure_3}(b), the results indicate near saturation at high carrier densities as reported in \cite{kubakaddi2010, Tuan2024, Scharf2012}, with acPE scattering being nearly the sole contributor compared to acDP. The total thermopower $S$ in Fig. \ref{fig:Figure_3}(c) reveals that $S_g$ primarily governs $S$, and at low densities $S$ decreases (increases) with larger (smaller) dielectric constants of the substrate, while saturating at high densities.

In Fig. \ref{fig:Figure_4}, we show the variation of the coefficients $S_d$, $S_g$, and $S$ with the interlayer distance $d$ between two BLG layers at $T = 10~\mathrm{K}$, $N_s = 10^{12}~\mathrm{cm^{-2}}$ for different dielectric substrates. The change of $S_d$ in Fig. \ref{fig:Figure_4}(a) shows saturation at large $d$, and acoustic deformation potential (acDP) gives the main contribution compared to piezoelectric (acPE) scattering. The change of $S_g$ in Fig. \ref{fig:Figure_4}(b) shows a similar trend with $S_d$ in that it saturates at large $d$, but in contrast, acPE contributes more to the acoustic phonons than acDP. For $S$ in Fig. \ref{fig:Figure_4}(c), which is the sum of $S_d$ and $S_g$, $S$ overlaps with $S_g$, increases at small $d$, and saturates at large $d$ ($d > 30~\text{\AA}$). Comparing the magnitudes of $S$ for different dielectric substrates shows that $S$ is larger for larger dielectric constants. More specifically, at $d = 10~\text{\AA}$, $T = 10~\mathrm{K}$, $N_s = 10^{12}~\mathrm{cm^{-2}}$: $S_{\mathrm{HfO_2}} \approx 968~\mu\mathrm{V/K}$, $S_{\mathrm{Al_2O_3}} \approx 611~\mu\mathrm{V/K}$, $S_{\mathrm{h\text{-}BN}} \approx 207~\mu\mathrm{V/K}$.

When the carrier densities in the two layers differ, the thermoelectric behavior changes in an interesting way. Specifically, in Fig. \ref{fig:Figure_5}, we show the variation of the three thermoelectric coefficients $S_d$, $S_g$, and $S$ as a function of density, where the carrier density in the first layer $N_{s1}$ varies from $0.5 \times 10^{12}~\mathrm{cm^{-2}}$ to $5.0 \times 10^{12}~\mathrm{cm^{-2}}$, while the carrier density in the second layer follows $N_{s2} = 2 \times N_{s1}$, at $T = 45~\mathrm{K}$, $d = 10 \times 10^{-8}~\mathrm{cm}$, for different dielectric environments. This is compared to Fig. \ref{fig:Figure_3}, where the two layers have equal densities varying with $N_s$. 

In Fig. \ref{fig:Figure_5}(a) compared to Fig. \ref{fig:Figure_3}(a) for acoustic phonon scattering, $S_d$ shows: $S_\mathrm{difference}^d < S_\mathrm{same}^d$ ($S_\mathrm{difference}^d > S_\mathrm{same}^d$) at low (high) densities. For example, at $N_{s1} = N_s = 0.5 \times 10^{12}~\mathrm{cm^{-2}}$, $N_{s2} = 2 \times N_{s1}$, $S_\mathrm{difference}^{\mathrm{HfO_2}} \approx 29.2~\mu\mathrm{V/K}$, $S_\mathrm{same}^{\mathrm{HfO_2}} \approx 35.7~\mu\mathrm{V/K}$; and at $N_{s1} = N_s = 5 \times 10^{12}~\mathrm{cm^{-2}}$, $N_{s2} = 2 \times N_{s1}$, $S_\mathrm{difference}^{\mathrm{HfO_2}} \approx 8.7~\mu\mathrm{V/K}$, $S_\mathrm{same}^{\mathrm{HfO_2}} \approx 3.7~\mu\mathrm{V/K}$.

Meanwhile, in Fig. \ref{fig:Figure_5}(b) compared to Fig. \ref{fig:Figure_3}(b), $S_g$ shows only slight differences under acoustic phonon scattering. For instance, at $N_{s1} = N_s = 0.5 \times 10^{12}~\mathrm{cm^{-2}}$, $N_{s2} = 2 \times N_{s1}$, the relative deviation is 
\[
\frac{S_\mathrm{same}^{\mathrm{HfO_2}} - S_\mathrm{difference}^{\mathrm{HfO_2}}}{S_\mathrm{same}^{\mathrm{HfO_2}}} \approx 10\%.
\]
Consequently, the total thermoelectric coefficient $S$ shown in Fig. \ref{fig:Figure_5}(c) and~\ref{fig:Figure_3}(c) are nearly identical because $S_g$ dominates the contribution to $S$.

When the interlayer distance increases to $d = 100\,\times\,10^{-8}\,\mathrm{cm}$, as shown in Fig. \ref{fig:Figure_6} with asymmetric carrier densities ($N_{s2} = 2\times N_{s1}$), the diffusion thermopower $S_d$ due to acoustic phonons (ac) shows negligible deviation across the three dielectric substrates compared to Fig. \ref{fig:Figure_5}(a), particularly at high densities (similar to what was observed in Fig. \ref{fig:Figure_4}(d). 

In contrast, the total thermoelectric coefficient $S$ increases for all three dielectric environments compared to the case with $d = 10\,\times\,10^{-8}\,\mathrm{cm}$ [Fig. \ref{fig:Figure_6}(c) vs. Fig. \ref{fig:Figure_5}(c)]. For example, at $N_{s1} = N_s = 0.5\times10^{12}\,\mathrm{cm^{-2}}$, $N_{s2} = 2\times N_{s1}$:
\[
S_{\mathrm{h\text{-}BN}}^{d = 100\times10^{-8}} \approx 3230~\mu\mathrm{V/K} > S_{\mathrm{h\text{-}BN}}^{d = 10\times10^{-8}} \approx 1091~\mu\mathrm{V/K};
\]
\[
S_{\mathrm{Al_2O_3}}^{d = 100\times10^{-8}} \approx 4775~\mu\mathrm{V/K} > S_{\mathrm{Al_2O_3}}^{d = 10\times10^{-8}} \approx 3362~\mu\mathrm{V/K};
\]
\[
S_{\mathrm{HfO_2}}^{d = 100\times10^{-8}} \approx 6097~\mu\mathrm{V/K} > S_{\mathrm{HfO_2}}^{d = 10\times10^{-8}} \approx 5326~\mu\mathrm{V/K}.
\]

On the other hand, by keeping the interlayer distance fixed at $d = 10 \times 10^{-8}~\mathrm{cm}$ and increasing the system temperature to $T = 90~\mathrm{K}$ for the BLG--GaAs--BLG structure with different dielectric substrates and unequal carrier densities ($N_{\mathrm{s2}} = 2 \times N_{\mathrm{s1}}$), as shown in Fig. \ref{fig:Figure_7}, we observe that $S_{\mathrm{d}}$ increases with temperature, particularly at low densities (compared to Fig. \ref{fig:Figure_5}(a), for all three dielectric environments. Specifically, for acoustic phonon scattering (ac), the diffusive thermopower values $S_{\mathrm{HfO_2}}^{\mathrm{d}}$, $S_{\mathrm{Al_2O_3}}^{\mathrm{d}}$, and $S_{\mathrm{h\text{-}BN}}^{\mathrm{d}}$ at $T = 90~\mathrm{K}$ are approximately $57~\mu\mathrm{V/K}$, $48~\mu\mathrm{V/K}$, and $25~\mu\mathrm{V/K}$, respectively, while the corresponding values at $T = 45~\mathrm{K}$ are about $29~\mu\mathrm{V/K}$, $23~\mu\mathrm{V/K}$, and $9~\mu\mathrm{V/K}$. Meanwhile, $S_{\mathrm{g}}$ appears to be nearly temperature-independent [as compared to Fig. \ref{fig:Figure_5}(b)].

More notably, from Figs. \ref{fig:Figure_5} through \ref{fig:Figure_7}, when the carrier densities in the two BLG layers vary unequally, the contribution of acoustic deformation potential (acDP) dominates in $S_{\mathrm{d}}$ over acPE, whereas acPE provides the major contribution to $S_{\mathrm{g}}$ over acDP. Furthermore, $S_{\mathrm{g}}$ remains the dominant component over $S_{\mathrm{d}}$ in the total thermopower $S$, consistent with the results previously observed in Figs. \ref{fig:Figure_2}--\ref{fig:Figure_4}, where the carrier densities in both layers were varied equally.

\section{CONCLUSION}
We have investigated the thermoelectric coefficient $S$ of the BLG--GaAs--BLG double-layer system with different dielectric substrates: h-BN, Al$_2$O$_3$, and HfO$_2$, as functions of temperature $T$, interlayer distance $d$, and carrier densities $N_s$ for both equal and unequal values in the two BLG layers. The second layer affects the first through the double-layer screening function $\left( \frac{W_{ii}(q, T)}{V_{ii}(q)} \right)^2$ and the different dielectric environments. 

For the diffusive thermopower $S_d$, the deformation potential scattering by acoustic phonons (acDP) provides the dominant contribution compared to piezoelectric scattering (acPE). In contrast, for the phonon-drag thermopower $S_g$, acPE is the main contributor over acDP. As the total thermopower $S$ is the sum of $S_d$ and $S_g$, its magnitude is mostly governed by the behavior of $S_g$, and it reaches higher values with higher substrate dielectric constants. 

When the carrier density of the second BLG layer is twice that of the first ($N_{s2} = 2 \times N_{s1}$, i.e., unequal densities), $S$ exhibits notable variations compared to the case of equal layer densities. Specifically, at low temperatures and small interlayer spacing, for acoustic phonon scattering, $S_\mathrm{difference}^{d} < S_\mathrm{same}^{d}$ at low density and $S_\mathrm{difference}^{d} > S_\mathrm{same}^{d}$ at high density, for all three dielectric substrates. Meanwhile, the difference in $S_g$ between equal and unequal densities remains negligible. As the interlayer distance $d$ increases, the total thermopower $S$ increases for all three dielectric substrates. Conversely, when $d$ is fixed and temperature increases, $S_d$ increases---especially at low density for all substrates---while $S_g$ remains nearly temperature-independent.

This study presents a nearly comprehensive theoretical analysis of the thermoelectric coefficient in BLG--GaAs--BLG heterostructures with commonly used dielectric substrates. It provides a useful reference for experimentalists to identify the contributions from $S_d$ or $S_g$, and from acDP or acPE in acoustic phonon scattering. Moreover, systems with higher dielectric constants yield greater thermopower $S$.

\section*{Conflict of Interest}
The authors declare that there is no conflict of interest regarding the publication of this paper.

\bibliography{references2}

\end{document}